\documentclass[10pt]{iopart}
\usepackage{latexsym,epsfig,graphicx,bbm}

\newcommand{\half}{\mbox{$\textstyle \frac{1}{2}$}}
\newcommand{\ket}[1]{\left | \, #1 \right \rangle}
\newcommand{\bra}[1]{\left \langle #1 \, \right |}

\newcommand{\bracket}[3]{\left\langle #1 \left| #2 \right| #3 \right\rangle}

\newcommand{\av}[1]{\langle #1\rangle}
\newcommand{\outprod}[2]{\ket{#1}\bra{#2}}

\newcommand{\vac}{\ket{\textrm{vac}}}

\newcommand{\eqr}[1]{Eq.~(\ref{#1})}
\newcommand{\fir}[1]{Fig.~\ref{#1}}
\newcommand{\secr}[1]{Sec.~\ref{#1}}

\begin{document}

\paper[Exact matrix product solutions in the Heisenberg picture
...]{Exact matrix product solutions in the Heisenberg picture of
an open quantum spin chain}

\author{S R Clark$^{1,2}$, J Prior$^{2,3,4}$, M J Hartmann$^{6}$, D Jaksch$^{2,1}$ and M B Plenio$^{3,5,7}$}
\address{$^1$Centre for Quantum Technologies, National University of
Singapore, 3 Science Drive 2, Singapore 117543}
\address{$^2$Clarendon Laboratory, University of Oxford, Parks
Road, Oxford OX1 3PU, United Kingdom}
\address{$^3$Institute for
Mathematical Sciences, Imperial College London, SW7 2PG, United
Kingdom}
\address{$^4$Departamento de F\'{\i}sica Aplicada, Universidad Polit\'ecnica de Cartagena,
Cartagena 30.202, Spain}
\address{$^5$QOLS, The Blackett
Laboratory, Imperial College London, SW7 2BW, United Kingdom}
\address{$^6$Technische Universit{\"a}t M{\"u}nchen, Physik
Department I, James Franck Str., 85748 Garching, Germany}
\address{$^7$Institut f{\"u}r Theoretische Physik, Albert-Einstein-Allee 11, Universit{\"a}t Ulm, D-89069 Ulm, Germany}

\date{\today}
\ead{s.clark@physics.ox.ac.uk}

\begin{abstract}
In recent work Hartmann {\em et al} [Phys. Rev. Lett. {\bf 102},
057202 (2009)] demonstrated that the classical simulation of the
dynamics of open 1D quantum systems with matrix product algorithms
can often be dramatically improved by performing time evolution in
the Heisenberg picture. For a closed system this was exemplified
by an exact matrix product operator solution of the time-evolved
creation operator of a quadratic fermi chain with a matrix
dimension of just two. In this work we show that this exact
solution can be significantly generalized to include the case of
an open quadratic fermi chain subjected to master equation
evolution with Lindblad operators that are linear in the fermionic
operators. Remarkably even in this open system the time-evolution
of operators continues to be described by matrix product operators
with the same fixed dimension as that required by the solution of
a coherent quadratic fermi chain for all times. Through the use of
matrix product algorithms the dynamical behaviour of operators in
this non-equilibrium open quantum system can be computed with a
cost that is linear in the system size. We present some simple
numerical examples which highlight how useful this might be for
the more detailed study of open system dynamics. Given that
Heisenberg picture simulations have been demonstrated to offer
significant accuracy improvements for other open systems that are
not exactly solvable our work also provides further insight into
how and why this advantage arises.
\end{abstract}


\maketitle

\section{Introduction}
Developing a more detailed understanding of the numerous
intriguing phenomena displayed by strongly correlated quantum
systems is one of the major theoretical challenges in physics
today. To meet this challenge a formidable arsenal of
non-perturbative, renormalization and numerical techniques have
been devised. The success of these approaches has for the most
part been routed in situations at or close to equilibrium, while
comparatively little is known about the physics of
strongly-correlated systems far-from-equilibrium. Yet
non-equilibrium systems are both ubiquitous and of significant
practical interest in physics. A typical example is where a finite
sized strongly-correlated quantum system is driven
far-from-equilibrium by introducing couplings to several different
macroscopic reservoirs, forming an open quantum
system~\cite{Breuer}. Under these circumstances both analytical
and numerical descriptions of the behaviour of the system become
highly non-trivial.

An immediate need to study open quantum systems is given by the
inevitable decoherence and dissipation present in any realistic
experimental realization of a strongly correlated quantum system.
Numerous examples of such experiments now exist ranging from
arrays of Josephson junctions~\cite{Fazio}, ultra-cold atoms in
optical lattice~\cite{Bloch,Lewenstein}, ion
traps~\cite{Porras,Retzker,Friedenauer} and arrays of coupled
microcavities~\cite{Hartmann06,Angelakis,Greentree}. Beyond this,
however, open systems are becoming increasingly relevant in
themselves as efforts are made to both understand, and also
potentially exploit, the beautiful and subtle interplay of the
coherent many-body dynamics and incoherent quantum processes they
possess. One example can be found in quantum information
processing~\cite{Nielsen} where the suppression of noise is
typically considered a prerequisite. Despite this it is found that
certain dissipative processes can in fact assist in the
preparation of highly entangled quantum
states~\cite{Plenio02,Diehl,Kraus}. More generally some of the
most common occurrences of non-equilibrium physics are in
transport problems~\cite{Datta} relevant to numerous systems
including quantum contacts~\cite{Agrait}, molecular
motors~\cite{Andrieux}, molecular junctions~\cite{Nitzan} and
other low dimensional heat conducting quantum
systems~\cite{Segal}. In addition to revealing a wealth of
non-equilibrium phenomena, including non-diffuse heat
transfer~\cite{Cahill} and negative differential
conductance~\cite{Benenti}, the presence of noise has been found,
contrary to expectations, to enhance transmission efficiency
through a dissipative quantum network~\cite{Mohseni,Plenio08}
where it has a beneficial influence thanks to its interplay with
destructive quantum interference and energy
mismatches~\cite{Caruso}. It is therefore of technological
relevance to better understand quantum mechanical effects in
driven dissipative strongly-correlated systems in order to exploit
them in achieving more robust and efficient energy transfer in
artificial structures~\cite{Caruso} and
nanomaterials~\cite{Benenti}. Finally open quantum systems present
a virtually unexplored landscape of non-equilibrium phases
transitions whose properties are likely to differ considerably
from conventional equilibrium transitions~\cite{Sachdev}.

In this work we shall adopt a master equation~\cite{Breuer}
description of an open quantum system. Our attention is focussed
on a specific class of open quantum systems, described in detail
in \secr{model}, which are governed by a quadratic spinless
fermionic Hamiltonian and coupled to baths described by
linear\footnote{Throughout this paper we use on occasion the
phrase {\em linear} as a shorthand for operators which are 1st
order in fermionic creation and annihilation operators, as in
\eqr{lindbladop}.} Lindblad operators. Only very recently this
class of open system was solved semi-analytically~\cite{Prosen1}
with this solution later providing strong evidence~\cite{Prosen2}
identifying, somewhat unexpectedly, a phase transition in the
far-from-equilibrium open XY spin chain with boundary pumping, but
no losses otherwise. Despite being a specialized type of system
its relevance is elevated by the fact that only a very limited
number of exactly solvable master equation models are known,
namely those involving a single particle, harmonic oscillator or
spin. Indeed the presence of a non-equilibrium transition in this
solvable model suggests that it may well come to represent a
paradigm for such phenomena analogous to the Ising model for
quantum phase transitions.

In this work we present an entirely complementary exact solution
of this system for the Heisenberg picture evolution of commonly
required observables by employing a matrix product
ansatz~\cite{Fannes,Rommer,Ostlund} for the operators, often
called a matrix product operator (MPO). This result is a
significant extension of the exact MPO solution presented
in~\cite{Hartmann09} for the purely coherent limit of the same
system. By using this approach our work provides, through the
formal structure of the MPO solution, further physical insight
into why this model is exactly solvable. The utility of our
solution, however, is only truly revealed when it is combined with
powerful matrix product based numerical methods, of which density
matrix renormalization group (DMRG)~\cite{White92,Schollwock}, and
more recently its quantum information inspired extension to
time-evolution~\cite{Vidal03,Vidal04,White04}, are leading
examples. These numerical methods enable the exact Heisenberg
picture MPO solution of the dynamical evolution of operators to be
computed under very general situations which are not easily
accessible otherwise. Given the rich properties of the
model~\cite{Prosen2} this in itself may be very useful. However,
perhaps of even greater importance, it also provides a significant
non-trivial example of where Heisenberg picture MPO numerics is
exact for an open system. It was shown~\cite{Hartmann09} recently
that in some cases it is much more efficient and accurate to
simulate open quantum systems in the Heisenberg picture and thus
the underlying numerical method used here can be readily applied
to more general interacting systems. In contrast to the equivalent
Schr{\"o}dinger picture MPO numerics~\cite{Zwolak,Verstraete}
where the study of entanglement has provided a crucial
understanding of its strengths and
limitation~\cite{Calabrese,Schuch,Perales,Venzl}, the merits of
the Heisenberg picture numerics for general situations is far less
clear. Our result provides further evidence in understanding when
rigourously exact or good approximate Heisenberg picture MPO
solutions may exist. We note also that promising results have been
found very recently in combining solutions of the Heisenberg
equations of motion with matrix product representations of states
for bosonic systems~\cite{Reslen}.

The structure of this paper is as follows. In \secr{model} we
describe the master equation and system we shall solve exactly and
introduce a particular spin chain model that our later numerical
calculations will focus on. Our solution exploits the MPO
formalism and so \secr{mpo_sec} describes all the necessary
details. We then show in \secr{coherent_evol} that the Heisenberg
picture solution of many operators for a closed coherent system
possesses an MPO form with a finite dimension, a fact which is
crucial for the exact solution to be numerically accessible. In
\secr{master_eq_construct} we introduce an ancilla construction
which reproduces the underlying master equation introduced in
\secr{model}. A crucial component of this construction is the
tracing out of ancillae and the effect of this on an MPO is
described in \secr{tracing_out}. We then combine these
observations in \secr{exact_sol} to demonstrate that an MPO
solution with a bounded dimension, identical to that of the purely
coherent case, exists for the open system considered. The
versatility of this result is highlighted in \secr{numerical}
where we numerically determine the MPO solutions for several
situations, including the approach to stationarity and a sudden
quench of the transverse field. Finally in \secr{conclusions} we
conclude and comment on future work.

\section{Model}
\label{model} In its most general form the physical system
considered in the work is a 1D system of spinless fermions
governed by a quadratic Hamiltonian which reads
\begin{eqnarray}
H_s &=& \sum_{ij}\left[c^\dagger_i {\bf a}_{ij}c_j + \half
c^\dagger_i{\bf b}_{ij}c^\dagger_j + \half c_i{\bf
b}_{ij}c_j\right], \label{quadratic_ham}
\end{eqnarray}
where $c_j^\dagger$ is fermionic creation operator for site $j$.
By demanding $H_s$ to be Hermitian we can choose $\bf a$ to be
real symmetric and $\bf b$ to be real antisymmetric matrices. To
describe an open fermi lattice we adopt the quantum master
equation approach leading to a Heisenberg picture evolution of an
operator $O(t)$ that is governed by the Lindblad master equation
of the form~\cite{Breuer} (using $\hbar = 1$ throughout)
\begin{eqnarray}
\frac{{\textrm d} O(t)}{{\textrm d} t} &=& \mathcal{H}\{O(t)\} +
\mathcal{L}\{O(t)\}, \label{mastereq}
\end{eqnarray}
where the Hamiltonian and Lindblad superoperators are
time-independent and defined as
\begin{eqnarray}
\mathcal{H}\{O(t)\} &=&  i[H_s,O(t)], \label{super_ham}\\
\mathcal{L}\{O(t)\} &=& \sum_{\gamma}\left(L^\dagger_{\gamma} O(t)
L_{\gamma} - \half L^\dagger_{\gamma} L_{\gamma}O(t) - \half
 O(t)L^\dagger_{\gamma} L_{\gamma}\right), \label{super_lind}
\end{eqnarray}
respectively. Here $L_{\gamma}$ are Lindblad operators specifying
the coupling of the system to a set of Markovian baths. We place a
restriction on the operators $L_{\gamma}$ that they are linear in
the fermionic creation and annihilation operators with the form
\begin{eqnarray}
L_{\gamma} &=& \sum_j \left(\ell_{\gamma j}\,c^\dagger_j +
l_{\gamma j}\,c_j\right), \label{lindbladop}
\end{eqnarray}
where $\ell_{\gamma j}$ and $l_{\gamma j}$ are complex
coefficients. A final constraint, which shall been seen in
\secr{master_eq_construct} to be essential to our result, is that
$O(t)$ must be an even ordered operator, so
$\mathbbm{P}O(t)\mathbbm{P}=O(t)$ where
$\mathbbm{P}=\prod_{j=1}^N(1-2c^\dagger_jc_j)$ is the parity
operator. Since parity is conserved by \eqr{mastereq} with
quadratic $H$ and linear $L_\gamma$'s we require only that the
initial operator $O(0)$ is even.

Very recently this class of open systems was solved
semi-analytically~\cite{Prosen1} by an entirely different approach
to that which will be described here. In~\cite{Prosen1} a
sophisticated method of constructing a Fock space of operators was
employed which maps the Liouvillian into a form which can be
diagonalized by a procedure analogous to the famous solution of
the XY Hamiltonian~\cite{Lieb}. This solution gives access to a
range of properties of the system including expectation values of
observables for the non-equilibrium stationary state and
excitations, as well as the spectrum of so-called
rapidities~\cite{Prosen1}. We shall exploit this solution later in
\secr{numerical} for testing the approach to stationarity in a
dynamical setting.

The fermionic model outlined has considerable freedom in the
non-locality of the terms in $H_s$ and the operators $L_\gamma$.
We shall consider a concrete example within this class of open
fermi systems composed of an XY spin chain with boundary pumping,
as depicted in \fir{spinchain}. As is well known the Jordan-Wigner
transformation
\begin{eqnarray}
c^\dagger_j &=&
\left(\prod_{k=1}^{j-1}\sigma^z_k\right)\sigma^-_j, \quad \quad
\textrm{and} \quad \quad c_j =
\left(\prod_{k=1}^{j-1}\sigma^z_k\right)\sigma^+_j, \label{jwt}
\end{eqnarray}
which relates spin ladder operators $\sigma^\pm$ to fermionic
creation and annihilation operators maps the XY spin-chain
Hamiltonian
\begin{eqnarray}
H_{xy} &=& \sum_{j=1}^N
J\left(\frac{1+\gamma}{2}\sigma^x_j\sigma^x_{j+1} +
\frac{1-\gamma}{2}\sigma^y_j\sigma^y_{j+1}\right) + B\sum_{j=1}^N
\sigma^z_j, \nonumber
\end{eqnarray}
directly to a spinless fermionic Hamiltonian~\cite{Lieb} of the
type in \eqr{quadratic_ham}. Here $J$ is the strength of the
nearest-neighbour spin coupling, $\gamma$ is the anisotropy, $B$
is a transverse magnetic field, and $\sigma^\alpha_j$ is the
$\alpha=\{x,y,z\}$ Pauli spin operator on the $j$th spin. The
boundary pumping is described by the set of Lindblad operators
\begin{eqnarray}
L_1 = \sqrt{\Gamma^L_+}\,\sigma^+_1, \quad L_3 =
\sqrt{\Gamma^R_+}\,\sigma^+_N, \nonumber \\
L_2 = \sqrt{\Gamma^L_-}\,\sigma^-_1, \quad L_4 =
\sqrt{\Gamma^R_-}\,\sigma^-_N. \nonumber
\end{eqnarray}
where $\Gamma^{L,R}_{+,-}$ are positive coupling constants. This
essentially models a system where the two ends of the spin chain
are coupled to separate thermal and magnetic baths. For an
uncoupled chain, where $J=0$, the ratios of the local bath
couplings $\Gamma^{L,R}_-/\Gamma^{L,R}_+ = \exp(-2B/T_{L,R})$ give
the temperature of the thermal state that the baths drive the
boundary spins to. This spin chain setup is not only of importance
to heat and spin transport problems~\cite{Datta} in 1D but also
strong numerical evidence suggests it possesses a non-equilibrium
phase transition as $B$ is varied~\cite{Prosen2}. Later in
\secr{numerical} we shall present some exact numerical results for
the dynamical behaviour of this system possible only through the
solution that we will now describe.

\begin{figure}
\begin{center}
\includegraphics[width=8cm]{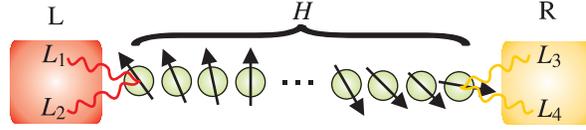}
\caption{A schematic plot of the open spin chain model considered
in this work relevant for transport problems. The coherent
evolution of the spin chain is described by an $XY$ type
Hamiltonian $H$ which maps to an effective quadratic fermionic
Hamiltonian. Boundaries of the chain are subject to couplings to
baths which are described by Lindblad operators $L_\gamma$ each of
which map to linear fermionic operators.}\label{spinchain}
\end{center}
\end{figure}

\section{Matrix product operators}
\label{mpo_sec} The framework in which we cast our exact solution
of \eqr{mastereq} is the matrix product representation of
operators. Given a system composed of $N$ sites each with a local
$d$-dimensional Hilbert space spanned by the states $\ket{j}$ we
define the tensor-product basis states as $\ket{{\bf j}} =
\ket{j_1}\ket{j_2}\dots\ket{j_N}$ where ${\bf j} =
(j_1,j_2,\dots,j_N)$ is a vector of physical indices. An arbitrary
operator $O$ acting on this system can then be expanded in the
operator basis $\outprod{{\bf j}}{{\bf k}}$ as $O = \sum_{{\bf
j},{\bf k}}o_{{\bf j},{\bf k}}\outprod{{\bf j}}{{\bf k}}$. A
matrix product operator (MPO) is where the coefficients $o_{{\bf
j},{\bf k}}$ of this expansion are expressed in the following
form~\cite{Rommer,Ostlund}
\begin{eqnarray}
o_{{\bf j},{\bf k}} &=& \bra{L}{\bf A}^{[1]j_1k_1}{\bf
A}^{[2]j_2k_2}\dots {\bf A}^{[N]j_Nk_N}\ket{R}, \label{mpocoeffs}
\end{eqnarray}
where ${\bf A}^{[n]j_nk_n}$ is a matrix, of dimension $\chi \times
\chi$, for each site $n$, selected by two independent physical
indices $j_n$ and $k_n$ for that site, while $\bra{L}$ and
$\ket{R}$ are $\chi$-dimensional row and column boundary vectors,
respectively. Each expansion coefficient $o_{{\bf j},{\bf k}}$ is
therefore encoded as a particular ordered product of $\bf A$
matrices associated to each site which is contracted to a scalar
by the fixed boundary vectors.

Given that there are in general exponentially many coefficients
$o_{{\bf j},{\bf k}}$ a matrix product representation in
\eqr{mpocoeffs} yields a highly compact description of an operator
if it requires only a small dimension $\chi$. For this reason, and
others, matrix product representations for both states and
operators have been applied with considerable success in a variety
of related numerical methods. The key to their success is that
many states or operators, for 1D systems at least, can be very
accurately approximated by a matrix product representation of
small dimension despite formally requiring a much larger
intractable dimension to be exact. In contrast to this the MPO
solutions we shall present require only a bounded dimension for
the representation to be exact when describing operators evolving
according to the open system introduced in \secr{model}. This
means that by utilizing one of these matrix product methods,
namely the time-evolving-block-decimation (TEBD) algorithm, we can
evaluate the exact solution numerically. However, beyond this much
is learnt about the nature of the solution by examining the
structure of the formal MPO solution itself. For this purpose we
utilize an entirely lower triangular form for all ${\bf
A}$-matrices, introduced
in~\cite{McCulloch1,McCulloch2,Crosswhite}, which permits exact
low-dimensional MPO representations for many operators to be
constructed easily. The key feature of this approach is that the
lower triangular form is preserved under the standard matrix
product manipulations such as direct sum or direct product. This
means that if an operator $O_A$ has a MPO representation with
matrices ${\bf A}$ of dimension $\chi_A$, and an operator $O_B$
has one with matrices ${\bf B}$ with dimension $\chi_B$, then the
operator $O_A + O_B$ has matrices ${\bf A}\oplus {\bf B}$ and
$O_AO_B$ has matrices ${\bf A}\otimes{\bf B}$ with a dimension of
at most $\chi_A + \chi_B$ and $\chi_A\chi_B$, respectively. Thus
much of the algebraic convenience of simple product operators
(i.e. $O = O_1\otimes O_2\otimes \cdots \otimes O_N$ over a system
of $N$ sites and is an MPO with a $\chi = 1$) can be extended to
highly non-trivial operators with an MPO dimension greater than
unity.

\begin{figure}
\begin{center}
\includegraphics[width=8cm]{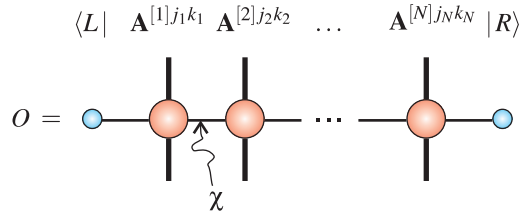}
\caption{The graphical representation of an MPO for an operator
$O$. For each lattice site $n$ the physical indices $j_n$ and
$k_n$, represented as the thick vertical lines respectively,
select a $\chi \times \chi$ matrix ${\bf A}^{[n]j_nk_n}$. The
joined up thin horizontal lines then represent the site-ordered
multiplication of these matrices, and finally the small circles at
the ends depict the boundary vectors $\bra{L}$ and $\ket{R}$
closing the chain.}\label{mpo}
\end{center}
\end{figure}

For our purposes we need only consider the simplest MPO with a
general $2 \times 2$ lower-triangular form. For an operator $O$ we
assign the following matrices to each site
\begin{eqnarray}
{\bf A} &=& \left[\begin{array}{cc}
     p & 0 \\
     q & r \end{array} \right].\nonumber
\end{eqnarray}
where $p$, $q$ and $r$ are $d \times d$ matrices representing
local operators on a site. Note that in this compact form of {\bf
A} the physical indices $j$ and $k$ are subsumed into physical
operators $p$, $q$ and $r$, while the row and column indices of
the ${\bf A}$ matrix are the internal $\chi = 2$ dimensional
indices of the MPO representation. To compute the full operator
described by assigning ${\bf A}$ to every site we note that the
standard multiplication of two ${\bf A}$ matrices is equivalent to
the tensor product of the physical operators they contain as
\begin{eqnarray}
{\bf A} \times {\bf A} &=& \left[\begin{array}{cc}
     p\otimes p & 0 \\
     q\otimes p + r\otimes q & r\otimes r \end{array} \right].\nonumber
\end{eqnarray}
For a longer string of multiplications ${\bf A} \times {\bf A}
\times \dots \times {\bf A}$ this generalizes to yield an operator
in the bottom left corner which is the sum of all terms of the
form $r \otimes\cdots\otimes r \otimes q \otimes p \otimes \cdots
\otimes p$ with the location of the $q$ operator in the string
translated. Finally for a lower-triangular MPO the full operator
is extracted via the left and right boundary states $\bra{L} =
(0,1)$ and $\ket{R} = (1,0)^T$, which select the bottom left
operator ``matrix element'' from the matrix product. Using
appropriate choices of $a$, $b$ and $c$ many useful
single-particle operators can be formed, for example $\sum_j
\sigma^z_j$ is formed by each site having a
matrix~\cite{McCulloch1}
\begin{eqnarray}
{\bf A} &=& \left[\begin{array}{cc}
     \mathbbm{1} & 0 \\
     \sigma^z & \mathbbm{1} \end{array} \right].\nonumber
\end{eqnarray}
Notice that an MPO representation is based on a tensor-product
structure and therefore implicitly assumes commutativity between
local operators appearing in the ${\bf A}$ matrices for different
lattice sites. The local operators cannot therefore be fermionic
directly. For products of such operator sums, which we shall
consider shortly, this means that MPO's always arrange the
resulting local operators in lattice site ordering.

\section{Exact MPO solution for a closed system}
\label{coherent_evol}Using the lower-triangular MPO formalism we
reexpress the finite-dimensional MPO solution described in
\cite{Hartmann09} for any fermionic operator governed by purely
coherent evolution with a quadratic $H_s$. To do this we need only
consider an arbitrary local sum of creation and annihilation
operators $C_\ell = x c_\ell + y c^\dagger_\ell$. The formal
solution to the equation of motion of this operator has the
standard form
\begin{eqnarray}
C_\ell(t) &=& e^{\mathcal{H} t}\{C_\ell\} = e^{i H_s t}C_\ell
e^{-i H_s t}. \nonumber
\end{eqnarray}
It can be readily shown that the action of $\mathcal{H}$ on
$C_\ell$ for a quadratic $H_s$ is
\begin{eqnarray}
\mathcal{H}\{C_\ell\} &=& i[H_s,C_\ell] \nonumber \\
&=& ix\sum_{j}\left\{{\bf a}_{j\ell}c_j + {\bf
b}_{j\ell}c^\dagger_j\right\} + iy\sum_{j}\left\{{\bf
a}_{j\ell}c^\dagger_j + {\bf b}_{j\ell}c_j\right\}, \nonumber
\end{eqnarray}
and thus $C_\ell$ is transformed into a sum of linear operators
spread across the lattice. The linearity of $\mathcal{H}$ implies
that its repeated application any integer number of times $p$ as
$\mathcal{H}^p\{C_\ell\}$ generates only a linear operator. Now
since the formal solution of the equation of motion can be
expanded as
\begin{equation}
C_\ell(t) =
\sum_{p=0}^\infty\frac{t^p}{p!}\mathcal{H}^p\{C_\ell\}, \nonumber
\end{equation}
we establish the well known fact that the Heisenberg picture
unitary time evolution of the operator $C_\ell$ governed by a
quadratic Hamiltonian is closed. The general solution can then be
written as
\begin{equation}
C_\ell(t) = \sum_{j=1}^N\left(\alpha_{\ell j}(t)c_j + \beta_{\ell
j}(t)c^\dagger_j\right),
\end{equation}
where $\alpha_{\ell j}(t)$ and $\beta_{\ell j}(t)$ are
time-dependent complex coefficients containing all the non-trivial
features of the evolution. To recast this solution in MPO form we
apply an inverse Jordan-Wigner transformation back to the
equivalent spin representation giving
\begin{equation}
\label{creationtime} C_\ell(t)
=\sum_{j=1}^N\left(\prod_{k=1}^{j-1}\sigma^z_k\right)\left(\alpha_{\ell
j}(t)\sigma^+_j + \beta_{\ell j}(t)\sigma^-_j\right).
\end{equation}
The spin operator on the righthand side can be expressed as a
simple $2 \times 2$ lower-triangular MPO, independent on the
number of sites $N$, with site-dependent matrices
\begin{eqnarray}
\label{c_operator}
{\bf A}^{[j]} &=& \left[\begin{array}{cc}
     \mathbbm{1}_j & 0  \\
     X_{\ell j}(t) & \sigma^z_j \end{array} \right],
\end{eqnarray}
where $X_{nj}(t) = \alpha_{\ell j}(t)\sigma^+_j + \beta_{\ell
j}(t)\sigma^-_j$. From our earlier discussion the bottom left
$X_{\ell j}(t)$ operator inserts the necessary site and time
dependent superposition of spin raising and lowering operators
into the product, while the bottom right $\sigma^z_j$ operator
creates the Jordan-Wigner operator string which establishes the
appropriate anti-commutative behaviour.

Since the evolution is unitary the solution to $c^\dagger_j(t)$
and $c_j(t)$ for all sites $j$ automatically provides the time
evolution for any string of local sums of creation and
annihilation operators, i.e. $(C_pC_q \dots C_k)(t) =
C_p(t)C_q(t)\dots C_k(t)$. Two consequences of this are that the
dynamics of a quadratic Hamiltonian $H$ conserves the order of any
initial fermionic operator and the MPO solution for the operator
string is simply the direct product of the MPO solution for each
constituent $C$-operator given in \eqr{c_operator}. The latter
then straightforwardly determines the fixed matrix product
dimension $\chi$ required for the solution of any given operator
string so $\chi = 2^n$ for an $n$th order operator. For example, a
general quadratic operator $C_p(t)C_q(t)$ has a $4 \times 4$ MPO
representation, independent of $p$ and $q$ given by matrices for
each site $j$ as
\begin{eqnarray}
\label{quadratic_sol}
{\bf B}^{[j]} &=& \left[\begin{array}{cccc}
     \mathbbm{1}_j & 0 & 0 & 0 \\
     X_{qj}(t) & \sigma^z_j & 0 & 0 \\
     X_{pj}(t) & 0 & \sigma^z_j & 0 \\
     X_{pj}(t)X_{qj}(t) & X_{pj}(t)\sigma^z_j & \sigma^z_j X_{qj}(t) & \mathbbm{1}_j\end{array}
     \right],
\end{eqnarray}
where $X_{pj}(t)$ and $X_{qj}(t)$ are the site-dependent $X$
operators associated to $C_p(t)$ and $C_q(t)$. The Kronecker
product of the MPO solutions gives the appropriately enlarged
boundary vectors $\bra{L} = (0,0,0,1)$ and $\ket{R} = (1,0,0,0)^T$
which select the accumulated operators in the bottom left corner
as
\begin{eqnarray}
C_p(t)C_q(t) &=& \sum_{j=1}^N X_{pj}(t)X_{qj}(t) +
\sum_{i=1}^N\sum_{j>i}
X_{pi}(t)\left(\prod_{k=i}^{j-1}\sigma^z_k\right)X_{qj}(t)
\nonumber \\
&& +
\sum_{i=1}^N\sum_{j<i}\left(\prod_{k=j}^{i-1}\sigma^z_k\right)
X_{qj}(t)X_{pi}(t) \nonumber.
\end{eqnarray}
A general feature of such solutions for strings of $C$ operators
is that each constituent $C$ operator contributes its own $X$
operator to the representation and shows how highly constrained
the evolution of operators is in the space of operators, a fact
which has ultimately permitted such a compact representation.

Common spin-chain observables such as $\sigma^z_j$,
$\sigma^x_j\sigma^x_{j+1}$, and $\sigma^y_j\sigma^y_{j+1}$ are
contained in this class of $4 \times 4$ MPO's. Long-range
correlations like $\sigma^z_p\sigma^z_q$ involve quartic fermionic
operators, independent of $p$ and $q$ and thus require $\chi=16$.
However, the behaviour of some operators can be very different.
For local spin observables such as $\sigma^x_j$ and $\sigma^y_j$
the fermionic representation obtained via an inverse Jordan-Wigner
transformation acquires a linearly growing order with the site
index $j$ due to the string of $(1-2c^\dagger c)$ operators which
appear. Such an operator could then require an exponentially
growing MPO dimension $\chi=2^{|j|}$ to describe its exact
solution. Correlations like $\sigma^y_p\sigma^y_q$ behave
similarly with an exponentially growing dimension $\chi=4^{|p-q|}$
dependent on their separation. What we shall now show in the
remainder of this paper is that the $\chi$ required for the MPO
solution of even ordered fermionic operators evolving according to
the open system described in \secr{model} is identical to that of
the purely coherent system.

\section{Ancilla master equation construction}
\label{master_eq_construct} Open quantum systems typically arise
when the system of interest interacts with a large bath or
reservoir, often identified as the system's environment. Using
this approach Lindblad master equations can be rigorously derived
using various microscopic models of the system-environment
interactions under the Born-Markov approximation and in the limit
of extremely large reservoirs~\cite{Breuer}. To prove that an
exact finite dimensional MPO representation exists for the open
systems introduced in \secr{model} we shall instead employ a
derivation of a master equation similar to that of non-selective
continuous measurement~\cite{Breuer}. While this construction
itself is perhaps less physically motivated it has the advantage
for our purposes that it yields a Lindblad master equation exactly
with no additional approximations.

A non-selective continuous measurement process involves dividing
time into small intervals of length $\delta t$ with each interval
associated to a separate independent ancilla (or probe) forming a
time-ordered chain. At the beginning of each interval $\delta t$
the system evolves coherently and interacts with the associated
ancilla which is subsequently measured at the end of the interval.
Depending on the interaction, measurement and ancilla initial
state this setup represents a general indirect continuous
monitoring of the system~\cite{Braginsky,Breuer}. In the case
where the indirect measurement is ideal the evolution of the
system is frozen by the quantum-Zeno effect. For more general
imperfect measurements the system evolves according to a master
equation with Hermitian Lindblad operators. In order to model the
linear fermionic Lindblad operators introduced in \eqr{lindbladop}
we modify this construction slightly by considering a different
class of system-ancilla coupling and trace out rather than measure
the ancilla at every time step. As we shall show below this setup,
depicted in \fir{ancilla_model}, produces in the continuous limit
an effective evolution of the system that is again described
exactly by a Markov master equation with the chain of ancillae
representing a manifestly delta-correlated environment in time.

\begin{figure}
\begin{center}
\includegraphics[width=8cm]{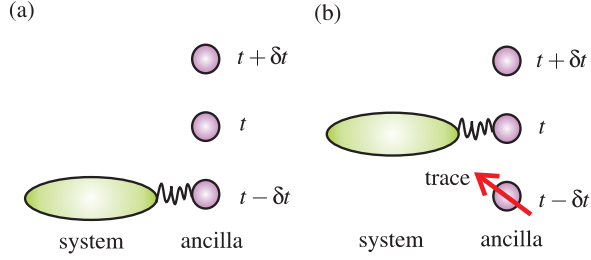}
\caption{(a) The ancilla construction can be visualized as a
time-ordered chain of ancilla systems associated to each interval
of time $\delta t$ and otherwise frozen in the initial state
$\ket{0}$ until that particular time. At time $t-\delta t$ the
ancilla-system Hamiltonian switches the interaction on with the
appropriate $t-\delta t$ ancilla for a duration $\delta t$. (b) At
time $t$ the Hamiltonian switches the interaction to the next
ancilla. All earlier ancilla never interact with the system again
and can be traced out.}\label{ancilla_model}
\end{center}
\end{figure}

The constructions begins by augmenting the system of $N$ sites
with a chain of $\tau+1$ ancilla sites described by the fermionic
modes $a_{\tt t}$ with ${\tt t}=0,1,\cdots,\tau$. Occupation
states of the system + ancillae are chosen to be defined by the
specific mode ordering
\begin{equation}
\ket{{\bf n},{\bf m}} =
(c_1^\dagger)^{n_1}\cdots(c_N^\dagger)^{n_N}(a_\tau^\dagger)^{m_\tau}\cdots(a_0^\dagger)^{m_0}\vac,
\label{occupancystate}
\end{equation}
where ${\bf n}=(n_1,\cdots,n_N)$ and ${\bf m}=(m_0,\cdots,m_\tau)$
are binary vectors of occupation numbers over the ancillae and
system modes, respectively. By placing the ancillae modes to the
right this choice, in conjunction with the Jordan-Wigner
transformation defined in \eqr{jwt}, ensures that any system
operator has a spin equivalent of the form\footnote{For notational
clarity we do not distinguish symbolically between a fermionic
operator and its Jordan-Wigner transformed spin equivalent. It
should be clear from the context which is implied.} $O_s(t) =
O_s(t)\otimes\mathbbm{1}_a$, where $\mathbbm{1}_a$ is the identity
over the corresponding ancillae spins. This enables the tracing of
spins to be completely equivalent to the tracing of the
corresponding fermionic mode. The ancilla mode label ${\tt t}$ is
essentially a time label denoting at which time interval the full
Hamiltonian of the system will involve that ancilla mode, as
depicted in \fir{ancilla_model}. In \eqr{occupancystate} we have
also ordered the ancillae amongst themselves so their time label
increasing inwards from the right so tracing can proceed
iteratively from the boundary. The full Hamiltonian of the system
+ ancillae is composed of two parts; the time-independent system
Hamiltonian $H_s$ involving only system modes, and $H_i(t)$ which
is a time-dependent interaction Hamiltonian between the system and
ancillae modes. The time-dependence of $H_i(t)$ is taken to be
piece-wise constant over intervals $\delta t$ giving a full
Hamiltonian
\begin{eqnarray}
H(t) &=& H_s + H_i(t), \label{interaction}  \\
&=& H_s + \sqrt{\frac{\kappa}{\delta t}}\sum_{{\tt t}=0}^\tau
\Theta({\tt t}\delta t - t) \Theta(t - {\tt t}\delta t-\delta t)
(a_{\tt t}^\dagger S + S^\dagger a_{\tt t}), \nonumber
\end{eqnarray}
where $S$ is a system operator and $\Theta(t)$ is the Heaviside
function. Notice that the interaction between the lattice and
ancilla in \eqr{interaction} depends on $\delta t$ and is singular
in the limit $\delta t \rightarrow 0$. This is physically required
in order for the ancilla to have a finite influence on the lattice
in the limit of a vanishingly small interaction
time~\cite{Braginsky}. Also the ancillae possess a zero
self-Hamiltonian so the only dynamics acting upon them is that
generated by the terms in $H_i(t)$. As a final definition for this
construction we take the initial time $t=0$ state $\rho$ of the
system + ancillae to have all ancillae modes unoccupied, but
otherwise arbitrary.

Let us focus on a particular time $t = T\delta t$. Between the
time $t$ and $t + \delta t$ the Hamiltonian $H(t)$ is time
independent and only involves the system modes and the ancilla
mode $a_T$. For this reason we shall, without loss of generality,
restrict our considerations to these modes only\footnote{Ancillae
modes $T+1,\cdots,\tau$ which are yet to interact are spectators
in the proceeding calculation since neither $O(t)$ nor $H(t)$
contain any of these modes. The ancillae modes $0,\cdots,T-1$
which have previously interacted may be contained in $O(t)$. The
proceeding calculation is the same regardless of whether these
modes are traced out before or after the considered time interval
$\delta t$. Thus for brevity we assume that they have been traced
out before in the same fashion as we shall trace out ancilla mode
$T$ below.}. To make the connection to MPOs transparent we perform
a Jordan-Wigner transformation and work with a spin
representation. The initial density matrix at time $t=0$ becomes a
spin state in which the lattice and all ancilla spins are
uncorrelated $\rho=\rho_s\otimes\outprod{\uparrow}{\uparrow}_T$
with $\rho_s$ being an arbitrary spin state of the system. At time
$t$ the initial operator $O(t)$ for the system + ancilla spin is
composed of system modes only and so it transforms to $O(t)=
O_s(t)\otimes\mathbbm{1}_T$, where $O_s(t)$ is the system operator
resulting from earlier evolution. Similarly the relevant
interaction terms in $H(t)$ transform to the spin operators
\begin{eqnarray}
H_{i}(t) &=& \sqrt{\frac{\kappa}{\delta t}}\left\{\sigma^-_T
\mathbbm{P}S + S^\dagger \mathbbm{P}\sigma^+_T\right\}, \nonumber
\end{eqnarray}
where $\mathbbm{P}=\prod_{j=1}^N\sigma^z_j$ is the spin equivalent
of the parity operator for the system. During the time interval
$\delta t$ the formal solution for the evolution is
\begin{equation}
O(t + \delta t) = e^{i \{H + H_i(t)\} \delta t} O(t) e^{-i \{H +
H_i(t)\} \delta t}, \label{formalsol}
\end{equation}
which will in general leave an operator defined over both the
lattice and ancilla spins. Since $\av{O(t+\delta t)} =
\tr_{as}[\rho O(t+\delta t)]$ the effective time-evolved lattice
operator is defined by tracing out ancilla spin $T$ resulting in
partial expectation value $O_s(t+\delta t) =
\bracket{\uparrow}{O(t+\delta t)}{\uparrow}_T$. The expectation
value $\av{O(t+\delta t)}$ can then be expressed solely in terms
of system operators as $\av{O(t+\delta t)} = \tr_{s}[\rho_s
O_s(t+\delta t)]$. Expanding \eqr{formalsol} to 2nd order gives
\begin{eqnarray}
O(t+\delta t) &=& O(t) + i\delta t[H_s+H_i(t),O(t)] \nonumber \\
&& \, + \frac{(i\delta t)^2}{2!}[H_s+H_i(t),[H_s+H_i(t),O(t)]] +
\cdots, \nonumber
\end{eqnarray}
which can be simplified considerably after the partial expectation
value is taken due to the special choice of interaction $H_i(t)$
and ancilla initial state $\ket{\uparrow}_T$. In particular
$\bracket{\uparrow}{H_i(t)}{\uparrow}_T= 0$ signifying that the
ancilla has no direct back action on the
system~\cite{Braginsky,Breuer}. The surviving 2nd order term
involving $H_i(t)$ is $[H_i(t),[H_i(t),O(t)]] = H_i(t)^2 O(t)
-2H_i(t) O(t) H_i(t) + O(t)H_i(t)^2$ which then also simplifies
since
\begin{eqnarray}
\bracket{\uparrow}{H_i(t)^2}{\uparrow} &=& \frac{\kappa}{\delta
t}S^\dagger S, \nonumber \\
\bracket{\uparrow}{H_i(t) O(t) H_i(t)}{\uparrow} &=&
\frac{\kappa}{\delta t}S^\dagger \mathbbm{P}O_s(t)\mathbbm{P} S,
\nonumber
\end{eqnarray}
Using the resulting evolution
\begin{eqnarray}
O_s(t+\delta t) &=& O_s(t) + i\delta t[H_s,O_s(t)] - \frac{\delta t^2}{2}[H_s,[H_s,O_s(t)]] \label{parity_evol} \\
&& \, + \frac{\kappa\delta t}{2}\left(2S^\dagger \mathbbm{P}
O_s(t) \mathbbm{P} S - S^\dagger SO_s(t) - O_s(t)S^\dagger
S\right) + \cdots, \nonumber
\end{eqnarray}
an equation of motion is formed by taking the continuous limit as
\begin{eqnarray}
\frac{\partial}{\partial t}O_s(t) &=& \lim_{\delta t \rightarrow 0} \frac{O_s(t+\delta t) - O_s(t)}{\delta t}, \nonumber \\
&=& i[H_s,O_s(t)] + \frac{\kappa}{2}\left(2L^\dagger O_s(t) L -
L^\dagger LO_s(t) - O_s(t)L^\dagger L\right). \nonumber
\end{eqnarray}
An implicit inverse Jordan-Wigner transformation back to spinless
fermions can be assumed whereupon we see that the construction has
yielded a standard Lindblad master
equation~\cite{Plenio98,Gardiner} with Lindblad operators $L =
\mathbbm{P}S$. The construction can be straightforwardly extended
to account for multiple Lindblad operators $L_\gamma$ by
introducing more ancillae and additional interactions of the form
in \eqr{interaction} at each time interval. It also admits the
option of having explicitly time-dependent Lindblad operators
$L_\gamma(t)$.

The presence of the $\mathbbm{P}$ operator relating the coupling
$S$ to the resulting Lindblad operator $L$ has important
consequences. Following our requirements outlined in \secr{model}
our aim is for this construction to model linear $L$ operators.
For even parity operators $O(0)$ the $\mathbbm{P}$ operator plays
no role making the coupling $S$ equivalent to the Lindblad
operator and therefore linear also. For the same choice of linear
coupling $S$ odd parity operators $O(0)$ instead evolve according
to a different Lindblad superoperator $\bar{\mathcal{L}}$ of the
form
\begin{eqnarray}
\bar{\mathcal{L}}\{O(t)\} &=& -\left(S^\dagger O(t) S + \half
S^\dagger SO(t) + \half
 O(t)S^\dagger S\right),
\end{eqnarray}
with a sign flip of the first term signifying that the Lindblad
operators $L$ are now higher order. Alternatively, for odd parity
operators to evolve with linear Lindblad operators $L$ the
interaction $S$ must instead include the parity and be higher
order.

\section{Tracing out ancilla within an MPO}
\label{tracing_out} A crucial step in the master equation
construction is the repeated tracing out of an initially
uncorrelated ancilla. We now detail the consequences this step has
on the resulting MPO representation of the system operator
$O_s(t)$ given an MPO of the full operator $O(t)$. Specifically,
let us take $O(t)$ as being represented by MPO matrices ${\bf
A}^{[j]}$ for each site $j$ of dimension $\chi$, and, without loss
of generality, take the ancilla to be the last site $j=N+1$.

Given an initial density matrix $\rho_s\otimes\rho_a$
\fir{trace_out} shows that the MPO representing the system
operator $O_s(t)$, satisfying $\tr_{sa}[O(t)
\rho_s\otimes\rho_a]=\tr_s[O_s(t)\rho_s]$, can be found by
contracting in isolation the single site ancilla density matrix
$\rho_a$ with the matrices ${\bf A}^{[N+1]}$ of $O(t)$. The
contribution of the ancilla to the remaining MPO is then reduced
to a matrix ${\bf T} = \sum_{jk}{\bf A}^{[N+1]jk}(\rho_a)_{jk}$
whose effect is simply to transform the right boundary vector as
${\bf T}\ket{R} = \ket{R'}$. The MPO for $O_s(t)$, defined only
over system sites $j=1,\cdots,N$, then retains the same set of
matrices ${\bf A}^{[j]}$ for those sites, but possesses the new
right boundary vector $\ket{R'}$. Thus, so as long as the initial
density matrix between the system and ancilla is uncorrelated, the
dimension of the MPO for $O_s(t)$ is identical to that of $O(t)$.
This conclusion can be readily seen to hold for any number of
initially uncorrelated ancilla located at the right edge of the
total system.

\begin{figure}
\begin{center}
\includegraphics[width=10cm]{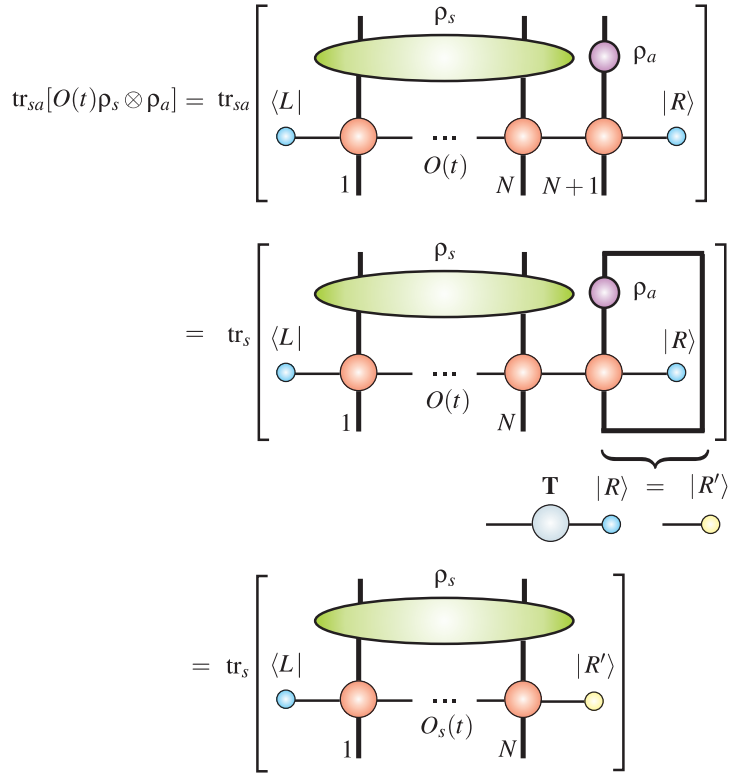}
\caption{Given that the full ancilla-system operator $O(t)$ is
described by an MPO the computation of the expectation value
$\av{O(t)}$ is found by contracting it with the system + ancilla
density matrix and performing a trace which contracts away in
pairs the remaining physical legs (vertical lines). Since the
system and ancilla are uncorrelated the trace over the ancilla can
be performed in isolation reducing its MPO matrices ${\bf
A}^{[N+1]}$ to a single matrix ${\bf T}$. The reduced system
operator $O_s(t)$ retains the same MPO matrices ${\bf A}^{[j]}$
spanning its sites $j=1,\dots,N$ but possess a new right boundary
vector $\ket{R'}$ resulting from transforming $\ket{R}$ with the
matrix ${\bf T}$. }\label{trace_out}
\end{center}
\end{figure}

For clarity let's consider what type of operators arise from using
an arbitrary left boundary vector with lower-triangular MPO's.
Using the earlier formal solution ${\bf B}^{[j]}$ for a generic
quadratic operator $C_p(t)C_q(t)$ given in \eqr{quadratic_sol},
the introduction of an arbitrary right boundary vector $\ket{R'} =
(\alpha,\beta,\gamma,\delta)^T$, while keeping $\bra{L}
=(0,0,0,1)$, gives a weighted sum of the bottom row of operator
``matrix elements''. More precisely it yields an operator
\begin{eqnarray}
G(t) &=& \delta\,\mathbbm{1} + \gamma\, \mathbbm{P}C_p(t) +\beta\,
 \mathbbm{P}C_q(t) + \alpha\, C_p(t)C_q(t), \label{admixture}
\end{eqnarray}
which now contains, modulo a parity operator $\mathbbm{P}$, linear
and zeroth order operators that are derived from the operators
appearing in the quadratic operator string with the admixture
determined by the components of $\ket{R'}$. In general an
arbitrary right boundary vector $\ket{R'}$ for an $n$th order
string $C_p(t)C_q(t)\dots C_k(t)$ generates a sum of all operators
of order less than or equal to $n$ derived from the constituents
of this string. When the parent string operator is an even parity
operator then lower order odd parity terms, such as the linear
terms in \eqr{admixture}, acquire an additional factor of
$\mathbbm{P}$. The opposite occurs when tracing a parent string
operator which has an odd parity.

The appearance of $\mathbbm{P}$ operators coincides with terms
which do not share the parity of the parent string
$C_p(t)C_q(t)\dots C_k(t)$ and thus they violate parity
conservation. These terms, however, are never generated by the
ancilla construction introduced, since tracing out the ancilla
spin in the specific initial state $\rho_a =
\outprod{\uparrow}{\uparrow}$ does not generate an arbitrary
boundary vector $\ket{R'}$. Instead the allowed structure for
$\ket{R'}$ can be readily discerned by again considering the
operator $O(t) = C_p(t)C_q(t)$ starting with the standard
$\ket{R}$. Tracing out the ancilla spin $j=N+1$ results in the
partial expectation value of the matrix ${\bf B}^{[N+1]}$ from
\eqr{quadratic_sol} with the state $\ket{\uparrow}$ and yields a
matrix $\bf T$ as
\begin{eqnarray}
{\bf T} &=& \left[\begin{array}{cccc}
     1 & 0 & 0 & 0 \\
     0 & 1 & 0 & 0 \\
     0 & 0 & 1 & 0 \\
     \zeta & 0 & 0 & 1\end{array}
     \right], \nonumber
\end{eqnarray}
where the complex number $\zeta =
\bracket{\uparrow}{X_{p}(t)X_{q}(t)}{\uparrow}$ is not necessarily
zero. Absorbing this matrix into the boundary gives $\ket{R'} =
(1,0,0,\zeta)^T$ signifying that only zeroth and quadratic terms
can appear. Depending on the nature of the system-ancilla
interaction tracing out a single ancilla spin for a general $n$th
order operator string can in principle generate a boundary vector
$\ket{R'}$ corresponding to a sum of operators of order $n, n-2,
n-4, \dots,{\textrm{mod}}(n,2)$. This is then consistent with the
resulting incoherent evolution preserving the parity of the
initial fermionic operators.

This shows that a given lower-triangular MPO's already possess the
capacity to describe a very specific class of mixed order
operators simply by varying one of the boundary vectors. As
described in \secr{coherent_evol} the coherent evolution according
to a quadratic Hamiltonian $H$ of any one $C$ operator in a string
is described by a specific time-dependent $X$ operator in its MPO
representation. This is true regardless of the boundary vector,
and so the same coherent quadratic evolution for this type of
mixed order operator is automatically captured by this MPO
solution.

\section{Exact open system MPO solution}
\label{exact_sol}

\subsection{Building an MPO solution}
The results of the preceding sections can be readily combined to
demonstrate that the bounded dimension of MPOs seen for coherent
quadratic Heisenberg picture evolution also applies to even parity
operators evolving according to the specific open system
introduced in \secr{model}. As mentioned in
\secr{master_eq_construct} for even parity operators the
requirement for linear Lindblad operators is met by using a linear
coupling operator $S$ between the system and ancilla. This, along
with a quadratic system Hamiltonian $H_s$, makes the full
time-dependent system + ancillae Hamiltonian $H(t)$ quadratic. If
all the ancillae modes are retained, as depicted in
\fir{trace_ancilla}(a), then the subsequent evolution would be
entirely coherent and would represent a {\em purification} of the
open dynamics of the system alone.

Since we have a coherent quadratic evolution, following the
discussion in \secr{coherent_evol}, an exact MPO solution of fixed
dimension therefore exists for the full operator $O(t)$. To
extract the reduced operator $O_s(t)$ for the system for any time
$0 \leq t \leq \tau\delta t$ the entire ancillae chain is traced
out, of which only those labelled up to $t$ have any relevance.
Given that the ancillae and system are uncorrelated initially the
tracing out of the ancillae has no effect on the resulting MPO
dimension for $O_s(t)$. The tracing out of the ancilla sites
yields a product of time-ordered ${\bf T}$ matrices\footnote{For
ancilla related to later times which have yet to interact the
transformation matrix ${\bf T} = \mathbbm{1}$.} as shown in
\fir{trace_ancilla}(b). The incoherent effects induced by the
ancilla are entirely captured by a time-dependent boundary vector
$\ket{R(t)} = {\bf T}_{t}\cdots{\bf T}_{\delta t}{\bf
T}_{0}\ket{R}$. This therefore establishes that for any
even-ordered initial system operator $O_s(0)$ there is an equality
of the required MPO dimension for its coherent evolution with a
quadratic Hamiltonian and its incoherent evolution with this
special type of open system. Since mixed order operators arising
from tracing out any single ancilla can also be coherently
evolved, with no change in their MPO dimension, this conclusion is
independent of when the tracing is performed. In particular
ancilla may be traced out immediately after they interact, as done
explicitly in \secr{master_eq_construct}, and thus the bounded MPO
dimension applies to the continuum limit as well. We demonstrate
this with some numerical examples in \secr{numerical}.

\begin{figure}
\begin{center}
\includegraphics[width=10.0cm]{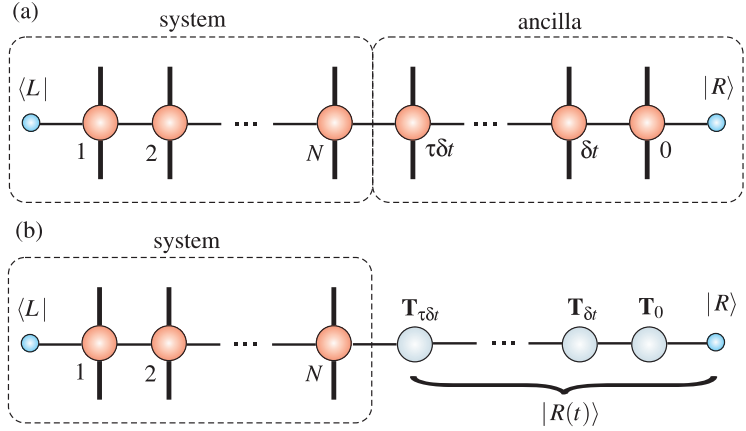}
\caption{A completely coherent representation of the incoherent
evolution of the open system introduced involves keeping track of
the full time-ordered chain of ancilla. If at some time $0\leq t
\leq \tau\delta t$ the reduced system operator is required then
the entire ancillae chain is traced out leaving behind a
time-ordered product of transformation matrices $\bf T$. The
effect of this is to leave the matrices ${\bf A}^{[j]}$ describing
the system sites unchanged from the form they acquired due to the
coherent evolution with the ancilla up to time $t$ and instead
introduces a time-dependent right boundary vector
$\ket{R(t)}$.}\label{trace_ancilla}
\end{center}
\end{figure}

\subsection{Properties of the MPO solution}
Here we make some additional comments on the MPO solution found.
Firstly, the existence of an exact solution for this open system
is not simply a consequence of the closure of the equations of
motion for the lowest order as it is for quadratic coherent
evolution in \secr{coherent_evol}. The lack of unitarity of the
evolution means that $(C_pC_q \dots C_k)(t) \neq C_p(t)C_q(t)\dots
C_k(t)$ in general so knowledge of the evolution of lower order
operators does not furnish us with knowledge of the evolution of
higher order products. Secondly, as shown in \secr{tracing_out}
the formal structure of the MPO solution with a time-varying
boundary vector with $\ket{R(t)} \neq (1,0,\cdots,0)^T$ implies
that the evolution of an initial $n$th order operator will involve
a special type of mixed order operator composed of equal or lower
order operators only. This behaviour for the operator order,
revealed by the formal MPO solution, is entirely consistent with
what is seen for other related open systems~\cite{Breuer}.

A classic example of this is provided by a modified version of the
damped harmonic oscillator. Here the coherent part of the master
equation in \eqr{super_ham} has $H = \omega b^\dagger b$ where
$\omega$ is the oscillator frequency and $b$ is its corresponding
bosonic annihilation operator. We then take the Lindblad
superoperator $\mathcal{L}$ in \eqr{super_lind} as being described
by a single linear Lindblad operator $L = \alpha b + \beta
b^\dagger$ analogous to the fermionic model introduced in
\secr{model}. Since any initial operator of the system can be
expanded as $O(t) = \sum_{nm} o_{nm}(t)(b^\dagger)^n(b)^m$ with
$n,m\geq 0$ and coefficients $o_{mn}(t)$, we need only consider
the effect of the righthand side of \eqr{mastereq} on a general
term within this expansion. The action of the coherent part is
simply $\mathcal{H}\{(b^\dagger)^n(b)^m\} =
i\omega(n-m)(b^\dagger)^n(b)^m$, while the Lindblad contribution
gives
\begin{eqnarray}
\mathcal{L}\{(b^\dagger)^n(b)^m\} &=&
\half(|\beta|^2-|\alpha|^2)(n+m)(b^\dagger)^n(b)^m \nonumber
\\
&& + |\beta|^2nm(b^\dagger)^{n-1}(b)^{m-1}
\nonumber \\
&& -\half\alpha^*\beta m(m-1)(b^\dagger)^n(b)^{m-2}
\nonumber \\
&& -\half\alpha\beta^* n(n-1)(b^\dagger)^{n-2}(b)^m. \nonumber
\end{eqnarray}
Thus whenever $|\beta|>0$ the action of
$(\mathcal{H}+\mathcal{L})$ is to leave the order of a constituent
term $(b^\dagger)^n(b)^m$ unchanged as $(n+m)$ or reduced by two.
The formal solution $O(t) =
\exp(\mathcal{H}+\mathcal{L})t\{O(0)\}$ for an initial operator
$O(0)$ of order $n$ (odd or even) will in general include
contributions only from orders $n, n-2, n-4,
\dots,{\textrm{mod}}(n,2)$. An identical type of analysis can be
performed for an open bosonic lattice defined by annihilation
operators $b_j$ for each site $j$ and again governed by a
quadratic Hamiltonian
\begin{eqnarray}
H &=& \sum_{ij}\left[b^\dagger_i {\bf \bar{a}}_{ij}b_j + \half
b^\dagger_i{\bf \bar{b}}_{ij}b^\dagger_j + \half b_i{\bf
\bar{b}}_{ij}b_j\right], \label{quadratic_bosonic_ham}
\end{eqnarray}
where ${\bf \bar{a}}$ and ${\bf \bar{b}}$ are real symmetric
matrices, along with linear Lindblad operators
\begin{eqnarray}
L_{\gamma} &=& \sum_j \left(\ell_{\gamma j}\,b^\dagger_j +
l_{\gamma j}\,b_j\right). \label{bosonic_lindbladop}
\end{eqnarray}
This readily confirms that the lack of growth of an operators
order seen for a single oscillator also applies to the fully
bosonic version of the model introduced in \secr{model}.

Applying a similar analysis on the fermionic lattice model itself
reveals that for the specific Lindblad superoperator $\mathcal{L}$
defined in \eqr{super_lind} only initially even ordered operators
display this closure property. In contrast odd ordered operators
can be shown to acquire a proliferating order under the repeated
application of $\mathcal{L}$. When such growth in the order occurs
the link between operator order and MPO dimension seen for the
coherent solution in~\secr{coherent_evol} suggest that the
dimension will not be bounded\footnote{Numerical evidence
following calculations like those to be presented
in~\secr{numerical} confirms this.}. Notice that our ancilla
construction applied to odd parity operators does not model
$\mathcal{L}$, but rather $\bar{\mathcal{L}}$. Incoherent
evolution according to $\bar{\mathcal{L}}$ reverses the situation
with odd parity operators now displaying no growth in their order.
Thus the ancilla construction presented in
\secr{master_eq_construct} models an incoherent evolution where
all operators $O(t)$ have a bounded order, which in turn permits
the bounded dimension MPO solution. It is only for even parity
operators, however, that this evolution corresponds to the precise
open system defined in \secr{model}.

Finally, the MPO solution offers a more efficient representation
than the closure of the operator order can provide on its own. In
particular once the exact $\chi$ is used for the MPO its
description only grows linearly with the number of sites $N$. This
is also an improved scaling compared to alterative approaches to
this open fermi system exploiting fermionic Gaussian
states~\cite{Bravyi}. They display a $N^2$ scaling and moreover
are restricted to considering initial states which are of Gaussian
form. The Heisenberg picture MPO approach used here can compute
properties for any initial state which can itself be well
approximated by a matrix product state.

\section{Numerical examples}
\label{numerical} Having shown that there exists a formal MPO
solution with a specific fixed dimension $\chi$ for a given system
operator we now show that this solution can be determined
numerically via Heisenberg picture evolution with the TEBD
algorithm~\cite{Vidal03,Vidal04,Hartmann09}. While the formal
solution presented has no restrictions regarding the locality of
the terms in $H_s$ and the Lindblad operators $L_\gamma$,
efficient integration of the equation of motion via the TEBD
algorithm requires that terms are nearest neighbour. Moreover
since linear Lindblad operators involving fermionic creation and
annihilation operators away from the boundaries acquire a
many-spin Jordan-Wigner $\sigma^z$ string the numerical solution
is restricted to noise terms on or one site in from the boundary.
This means that specific open XY spin chain model introduced in
\secr{model} can be solved with this numerical method.

\begin{figure}[t]
\begin{center}
\includegraphics[width=12cm]{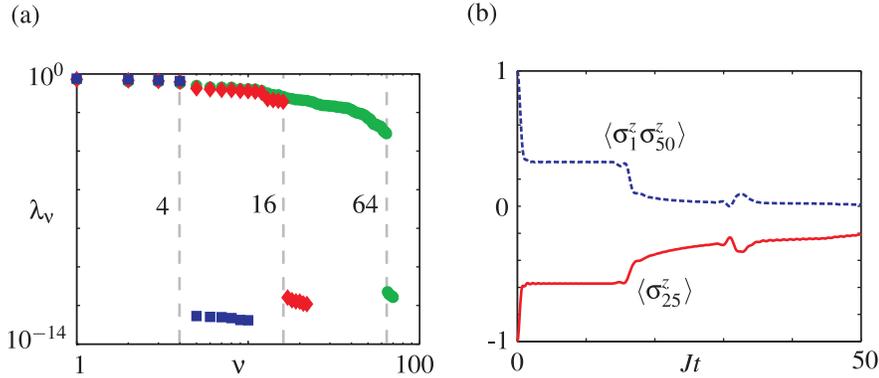}
\caption{(a) The Schmidt coefficients $\lambda_\nu$ for the
splitting between sites 25 and 26 are plotted on a log-log scale
for the MPO's of the operators $\sigma^z_{25}$,
$\sigma^z_1\sigma^z_{50}$ and $\sigma^x_{24}\sigma^y_{27}$ after a
time $Jt = 50$ of evolution. The analytical limit for the MPO
dimension for each operator is $\chi=4,16,64$, respectively and
these are depicted by the corresponding dashed vertical lines. (b)
The evolution of the central $z$-magnetization
$\av{\sigma^z_{25}}$ (solid) and boundary $z$-correlation
$\av{\sigma^z_1\sigma^z_{50}}$ (dashed) up to a time $Jt = 50$
starting from a spin-polarized initial state. All these
calculations are performed on an XY chain of length $N=50$ with
$\gamma = 0.75$, $B/J=1$, $\Gamma^L_-/J = 0.5$, $\Gamma^L_+/J =
0.3$, $\Gamma^R_-/J = 0.5$ and $\Gamma^R_+/J =
0.7$.}\label{fig:operator_plots}
\end{center}
\end{figure}

The numerical solution determined by TEBD unmistakably
demonstrates the existence of the bounded MPO dimension proven
above for this open system, just as it does for the coherent
limit~\cite{Prosen07}. When normalized according to the Frobenius
norm, where $\sum_{{\bf j},{\bf k}}o_{{\bf j},{\bf k}} = 1$, the
MPO solution produced by TEBD\footnote{The numerical MPO solution
will not be lower-triangular. Instead it will be a gauge
equivalent canonical solution which maintains an orthonormal
matrix product structure essential for stability and convergence
of the numerical algorithm.} is in canonical
form~\cite{Vidal03,PerezGarcia}
\begin{eqnarray}
o_{{\bf j},{\bf k}} &=& \bra{\Lambda^{[1]}}{\bf
\Gamma}^{[1]j_1k_1}{\bf \Lambda}^{[2]}{\bf \Gamma}^{[2]j_2k_2}
\dots {\bf \Lambda}^{[N]}{\bf
\Gamma}^{[N]j_Nk_N}\ket{\Lambda^{[N+1]}},
\end{eqnarray}
where the Schmidt decomposition~\cite{Nielsen} of the operator for
any contiguous bipartition is explicitly contained in the
representation~\cite{Zwolak}. Here ${\bf \Gamma}^{[n]j_nk_n}$ are
sets of matrices, different to the lower triangular matrices ${\bf
A}^{[n]j_nk_n}$ used earlier but performing the same function. The
new important addition to this representation are the diagonal
matrices ${\bf \Lambda}^{[n]}$ for the bulk (i.e. $n=2,\dots,N-1$)
with diagonal elements equal to the Schmidt coefficients
$\lambda_\nu$ of a bipartition after site $n$. The appropriately
sized boundary vectors are now $\bra{\Lambda^{[1]}}=(1,0,\dots,0)$
and $\ket{\Lambda^{[N+1]}} = (1,0,\dots,0)^T$, representing the
single unit Schmidt coefficient before site $1$ and after site
$N$. Once in this canonical form Schmidt coefficients allow the
effective MPO dimension of the operators to be identified by
counting the number of significant Schmidt coefficients $\epsilon
\leq \lambda_\nu \leq 1$, where $\epsilon$ is a some small
threshold.

For an XY chain of length $N=50$ (see \fir{fig:operator_plots} for
the parameters used) we have calculated the evolution of the
operators $\sigma^z_{25}$, $\sigma^z_1\sigma^z_{50}$ and
$\sigma^x_{24}\sigma^y_{27}$. In \fir{fig:operator_plots}(a) the
central Schmidt coefficients for the chosen operators after a time
$Jt = 50$ of evolution are shown. A clear cut-off in the
$\lambda_\nu$'s is seen where their value drops in excess of 11
orders of magnitude. This cut-off is robust to time-evolution and
the insignificant $\lambda_\nu$'s are numerical noise that may be
safely truncated away. The effective MPO dimension given by this
cut-off coincides with the dimension expected from the formal
solution. We may therefore rigidly enforce the exact MPO dimension
required and given the lack of truncation the only error in the
time integration comes from the customary Trotter expansion used
in TEBD. In \fir{fig:operator_plots}(b) the resulting
time-evolution of the central $z$-magnetization $\sigma^z_{25}$
and boundary-boundary $z$ correlation $\sigma^z_1\sigma^z_{50}$ is
shown for an initial spin-polarized state
$\ket{\downarrow\downarrow\cdots\downarrow}$. The transient
evolution displays plateaus caused by the time it takes for the
influence of the boundary pumping to propagate across the chain.
This is better illustrated in \fir{fig:stat_mag_profile} where the
evolution of the $z$-magnetization profile of the entire chain is
plotted up to a time $Jt=500$. Being plotted with a logarithmic
timescale it is apparent that the majority of the
$z$-magnetization in the bulk is eroded rapidly by the dynamics
from its initial value. However, in \fir{fig:quench}(a) a more
detailed comparison of the $z$-magnetization profile at a time
$Jt=500$ and the stationary profile~\cite{Prosen1} reveals that
for $N=50$ spins this time is only sufficient to drive the
boundary $z$-magnetization to their stationary values and that the
bulk is still far from stationary. Tests reveal that a
significantly longer evolution time is needed to achieve
convergence of the bulk $z$-magnetizations.

\begin{figure}[t]
\begin{center}
\includegraphics[width=6.75cm]{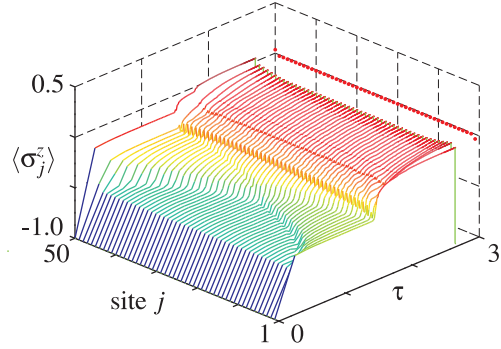}
\caption{The $z$-magnetization profile for the entire chain over
time plotted against $\tau = \log_{10}(1+Jt)$ up to a time of $Jt
= 500$. For visual convenience the stationary $t \rightarrow
\infty$ magnetization profile determined from the exact solution
given in \cite{Prosen1} is plotted at $\tau = 3$. See also
\fir{fig:quench}(a) for a comparison of the stationary profile
with that attained after a time $Jt = 500$. The Hamiltonian
parameters used are identical to those stated in
\fir{fig:operator_plots}. }\label{fig:stat_mag_profile}
\end{center}
\end{figure}

To demonstrate a time-dependent dynamical scenario\footnote{The
application of the TEBD method can be readily adapted to deal with
time-dependence in the Heisenberg picture.} we consider the
simplest case of an abrupt quench of the transverse field.
Specifically we evolve the initial spin-polarized state with $B/J
= 10$ for a time $Jt = 500$, analogous to the previous example.
Then at the time $Jt = 500$ the transverse field is switched
instantaneously to $B/J = 1$ and the evolution is continued. In
\fir{fig:quench}(b) the evolution of the central $z$-magnetization
$\av{\sigma^z_{25}}$ is shown as a function of time around the
quench point. For times $Jt < 500$ we see that there is a slow
change in $\av{\sigma^z_{25}}$ and it is still quite far from its
stationary value of $\av{\sigma^z_{25}}_s = -0.0161$. In the same
region of time \fir{fig:quench}(b) shows (dashed line) the
evolution of $\av{\sigma^z_{25}}$ from the previous example where
$B/J=1$ throughout which is slightly closer to its stationary
value of $\av{\sigma^z_{25}}_s = -0.0391$ but displays a similar
rate of convergence. For time $Jt > 500$, after the quench, there
is initially a rapid change in $\av{\sigma^z_{25}}$ which after a
time of approximately $15/J$ then settles down with small
oscillates around a new value which again differs from the
stationary value of the new transverse field. Instead this newly
acquired $z$-magnetization is very close to the non-stationary
value obtained via constant evolution with $B/J=1$. This shows
that even after a comparatively long evolution time the system has
retained a significant memory of its initial spin polarized state.

\begin{figure}[t]
\begin{center}
\includegraphics[width=12cm]{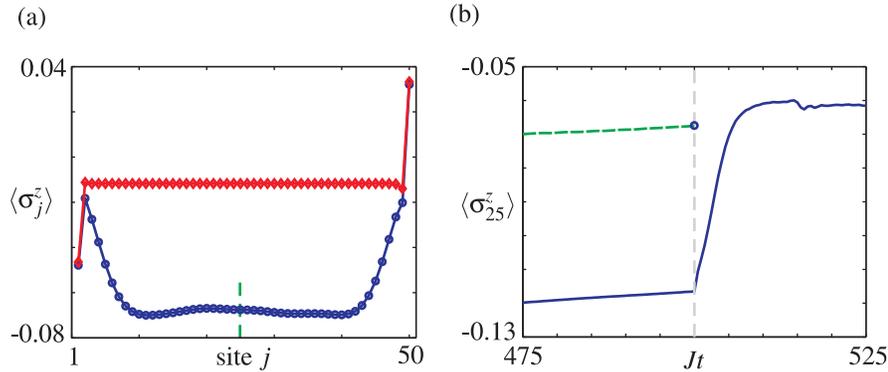}
\caption{(a) A comparison of the stationary ($\diamond$)
$z$-magnetization profile $\av{\sigma^z_j}$ with that attained
from a spin-polarized initial state ($\circ$) after evolving for a
time $Jt = 500$. The central $z$-magnetization
$\av{\sigma^z_{25}}$ is highlighted with a dashed line. (b) For a
sudden quench of the traverse field from $B/J=10$ to $B/J=1$ at a
time $Jt = 500$ the time evolution of the central
$z$-magnetization $\av{\sigma^z_{25}}$ is plotted. The dashed line
shows the evolution of $\av{\sigma^z_{25}}$ up to a time $Jt =
500$, already displayed in (a) and in \fir{fig:stat_mag_profile},
for a constant transverse field $B/J=1$. Aside from those stated
all other Hamiltonian parameters are identical to those in
\fir{fig:operator_plots}.}\label{fig:quench}
\end{center}
\end{figure}

\section{Conclusions}
\label{conclusions} We have presented a detailed study of the MPO
description of a specific class of open quantum systems governed
by a master equation with a quadratic spinless fermionic
Hamiltonian and linear fermionic Lindblad operators. By mapping
this master equation to an entirely coherent quadratic evolution
involving additional ancillae we have shown that the MPO
representation for the evolution of operators with even parity
possesses a finite and fixed dimension. This has revealed the
quadratic nature of the evolution underlying this class of master
equations and our ancilla construction gives decisive insight into
why it is exactly solvable. The formal structure of the MPO
representation also indicates how a given initial operator can
evolve into a specific type of mixed order operator, consistent
with behaviour seen in other simpler open systems. Exploiting the
fixed MPO dimension the TEBD algorithm allows the dynamical
evolution of operators in this non-equilibrium open quantum system
to be computed with a cost that is linear in the system size. The
dynamical behaviour accessible via the MPO solution presented
therefore complements the existing exact solution for this models
stationary states and spectral properties~\cite{Prosen1}. We have
exemplified this by computing some examples involving the approach
to stationarity and the response of the $z$-magnetization to a
sudden quench in the transverse field.

An interesting calculation, beyond the scope of the current work,
is to perform a dynamical quenching through the non-equilibrium
quantum phase transition. Such a dynamical calculation appear to
be very demanding with the Schr{\"o}dinger picture~\cite{Prosen2}.
The non-equilibrium transition manifests itself as a discontinuous
change in the $\av{\sigma^z_i\sigma^z_j} -
\av{\sigma^z_i}\av{\sigma^z_j}$ correlations, but not in other
local observables such as energy and magnetization. From the MPO
perspective of this work the behaviour of this transition appears
to be very reminiscent of the matrix-product type equilibrium
quantum phase transitions~\cite{Wolf}. Computing a dynamical
crossing of this non-equilibrium transition could help determine
the realistic adiabacity requirements for its observation.

Beyond this our work has provided an important and non-trivial
class of open systems with an exact Heisenberg picture MPO
representation. This may yet aid in determining other models where
such solutions exist. For instance it remains to be seen whether
Heisenberg picture simulability is readily related to the
integrability of the underlying model~\cite{Prosen07}. For example
a finite sized XXZ chain can be made integrable with appropriate
boundary fields, however it is not clear that an efficient
representation exists for commonly required local observables like
$\sigma^z$. This raises the question as to whether finite-sized
MPO representations of certain types of operators are possible for
systems possessing a Bethe-ansatz solution. This is an interesting
open problem and would reveal if the MPO formalism can aid in
evaluating otherwise very complicated quantities from these
solutions. For the presently studied XY model the non-interacting
nature of the effective fermi system for both the open and closed
system appears to be a crucial property permitting simulability,
which is more constraining than integrability alone.

Finally the MPO solution introduced may allow a better
understanding of the trade-off between efficiencies possible by
changing pictures. Future work~\cite{Prior} will look at how
quickly the accuracy of Heisenberg picture simulations breakdown
when they are applied to models which are only weakly perturbed
from the exact solution presented here. In the context of spin
chains the most obvious extensions outside the exact solution
would be additional $\sigma^z_j\sigma^z_{j+1}$ interaction terms
and/or dephasing noise.
\\
\\
SRC acknowledges very helpful correspondence with Toma\v{z} Prosen
on his exact solution of the model considered in this work, and
Ian McCulloch on matrix product operator formulations. SRC also
thanks the National Research Foundation and the Ministry of
Education of Singapore for support. DJ acknowledges support from
the ESF program EuroQUAM (EPSRC grant EP/E041612/1), the EPSRC
(UK) through the QIP IRC (GR/S82176/01), and the European
Commission under the Marie Curie programme through QIPEST. MBP
acknowledges support from the EPSRC (UK) grant EP/E058256/1, the
EU Integrated Project QAP supported by the IST directorate as
contract number 015848, the EU STREP HIP, a Royal Society Wolfson
Research Merit Award, and the EU STREP project CORNER. MJH
acknowledges support from the DFG via the Emmy Noether project
HA5593/1-1. Finally, JP was supported by the Fundaci\'on S\'eneca
grant 05570/PD/07 and Ministerio de Ciencia e Innovaci\'on project
number FIS2009-13483-C02-02.

\end{document}